\documentclass{ceab}   

\usepackage{epsfig}     
\usepackage{graphicx}   

\usepackage{ceabbib}     
\usepackage[T1]{fontenc}

\begin{document}
\countdef\pageno=0
\pageno=1

\def\tit{Correlation between solar magnetic field and brightness temperature}
\def\aut{I. SKOKI\'C et al.}
\def\str{1--11}

\title{Correlation between the solar magnetic field strength and the millimeter brightness temperature}

\author{I. SKOKI\'{C}\thanks{ivica.skokic@gmail.com}
, D. SUDAR and R. BRAJ\v{S}A\vspace{2mm}\\
\it Hvar Observatory, Faculty of Geodesy, University of Zagreb,\\ 
\it Ka\v{c}i\'{c}eva 26, HR--10000 Zagreb, Croatia
}

\maketitle
\def\even{Solar magnetic field and brightness temperature}

\begin{abstract}
Images of the Sun at millimeter wavelengths obtained by ALMA show a significant correspondence with the magnetograms. In this paper, we investigate this correspondence by comparing ALMA full-disk solar image taken at 1.2 mm with a SDO/HMI magnetogram and analyze their correlation. It is found that chromospheric network and active regions show a positive correlation where brightness temperature is increasing with the line-of-sight magnetic field strength, while sunspots have a negative correlation. Quiet Sun regions do not show any dependence of the brightness temperature with the magnetic field. Thermal bremsstrahlung is given as the best explanation for the observed correlations.
\end{abstract}


\section{Introduction}

Solar radiation in the (sub)millimeter wavelength range is mostly generated by free--free (thermal bremsstrahlung) emission from the interaction of free electrons with ions within the chromosphere, while in regions with strong magnetic fields, such as sunspots, a gyroresonant emission coming from the electrons spiraling along the magnetic field lines also plays an important role (see, e.g., \citealt{Wedemeyer2016review, Nindos2020review, Alissandrakis2020review}). Theoretical models predict that magnetic field modulates free--free opacity differently for ordinary and extraordinary wave modes, which should be measurable as somewhat lower and higher intensity from the non-magnetic level, when observing in circularly polarized radiation \citep{Loukitcheva2020review}. For gyroresonant emission, however, models predict that very strong magnetic fields are neccessary to have significant contribution to total radiation in the (sub)millimeter range, much stronger than those observed \citep{Brajsa2009}. 

Solar observations with the Atacama Large Millimeter/submillimeter Array (ALMA) have provided us with a new tool to study the Sun at (sub)millimeter wavelength range \citep{Bastian2018}. It has already produced new and interesting results (e.g. \citealt{Iwai2017, Brajsa2018,Nindos2018,Selhorst2019,Rodger2019,Loukitcheva2019,Sudar2019CEAB}). 
However, ALMA is not yet capable of providing circular polarization measurements of the Sun.

Various features in solar ALMA images have been identified to have a good correspondence with known solar structures. \citet{Brajsa2018} have reported that, at 1.2 and 3 mm, sunspots and inversion lines are darker, active regions and coronal bright points are brighter, while coronal holes and prominences have very small contrast from the quiet Sun background. They also found a very good correspondence with the \textit{Solar Dynamics Observatory} (SDO; \citealt{Pesnell2012SDO}) \textit{Atmospheric Imaging Assembly} (AIA; \citealt{Lemen2012SoPh}) 304~\r{A} and 1600~\r{A}  channels, and \textit{Helioseismic and Magnetic Imager} (HMI; \citealt{Schou2012HMI}) magnetograms.

The main goal of this paper is to analyze and quantize this good correspondence of ALMA solar images with HMI magnetograms indicating a close relationship between the observed millimeter brightness temperature and the magnetic field strength.

\section{Data and method}

We used a single dish image of the Sun from the ALMA Solar Commissioning and Science Verification Campaign (CSV), taken on December 18, 2015 at 20:12 UT \citep{White2017}. This image was selected because it was taken when the Sun was still very active and it contains examples of many different solar regions such as active regions, quiet Sun network and inter-network regions, sunspots, etc. Moreover, this data is well known and has been analyzed by many researchers. The image was made by fast scanning the Sun using a double circle pattern with a 12 m PM antenna in the band 6, in the spectral window 3 with a rest frequency of 248 GHz or equivalent wavelength of 1.2 mm. Limb brightening was removed from the ALMA image by the method presented in \citet{Sudar2019} and the image was scaled to the quiet Sun level of 5900 K, as suggested by \citet{White2017}. 

\begin{figure}[ht]
  \centering
   \epsfig{file=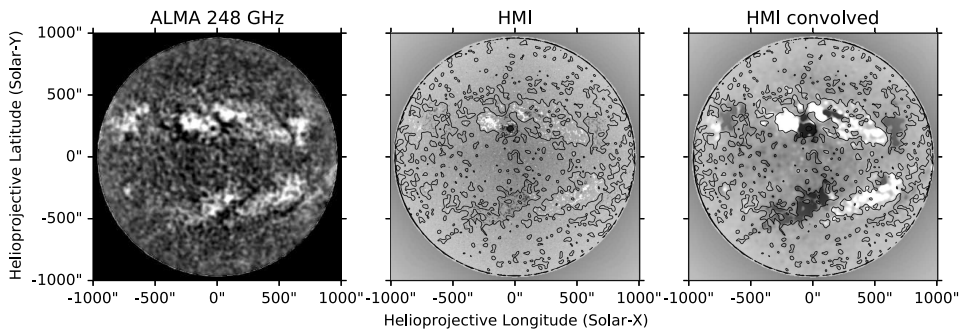,width=12.6cm}%
  \caption{HMI magnetogram resized (center image) to ALMA image (left) size and convolved with the ALMA beam (right). ALMA contours of 6100 K were overlaid on both HMI images. For better contrast, HMI and convolved HMI images were normalized to $\pm 100$ G and $\pm 20$ G, respectively.}
  \label{fig:alma_hmi_overview}
\end{figure}

For comparison with the ALMA image, a line-of-sight (LOS) magnetogram from the HMI instrument aboard SDO was used. The HMI image was first convolved with a Gaussian kernel with a standard deviation of $\sigma = \textrm{FWHM}/(2\sqrt{2\ln 2})\approx \textrm{FWHM}/2.355$, where FWHM is the Full Width at Half Maximum of the ALMA band 6 beam of the 12 m antenna (equal to 26.7 arcsec, which is also the effective spatial resolution). After that, the 4096x4096 pixel HMI image was resized to match 800x800 pixel ALMA image with an image scale of 3 arcsec per pixel. The ALMA image, the resized HMI image, and the convolved and resized HMI image are shown in Fig. \ref{fig:alma_hmi_overview}, in the left, middle and right panels, respectively. ALMA contours of 6100 K were overlaid on both HMI images for easier comparison with the ALMA image. A good correspondence can be seen between the bright ALMA regions and the HMI areas with a large magnetic field intensity.

\begin{figure}[ht]
  \centering
   \epsfig{file=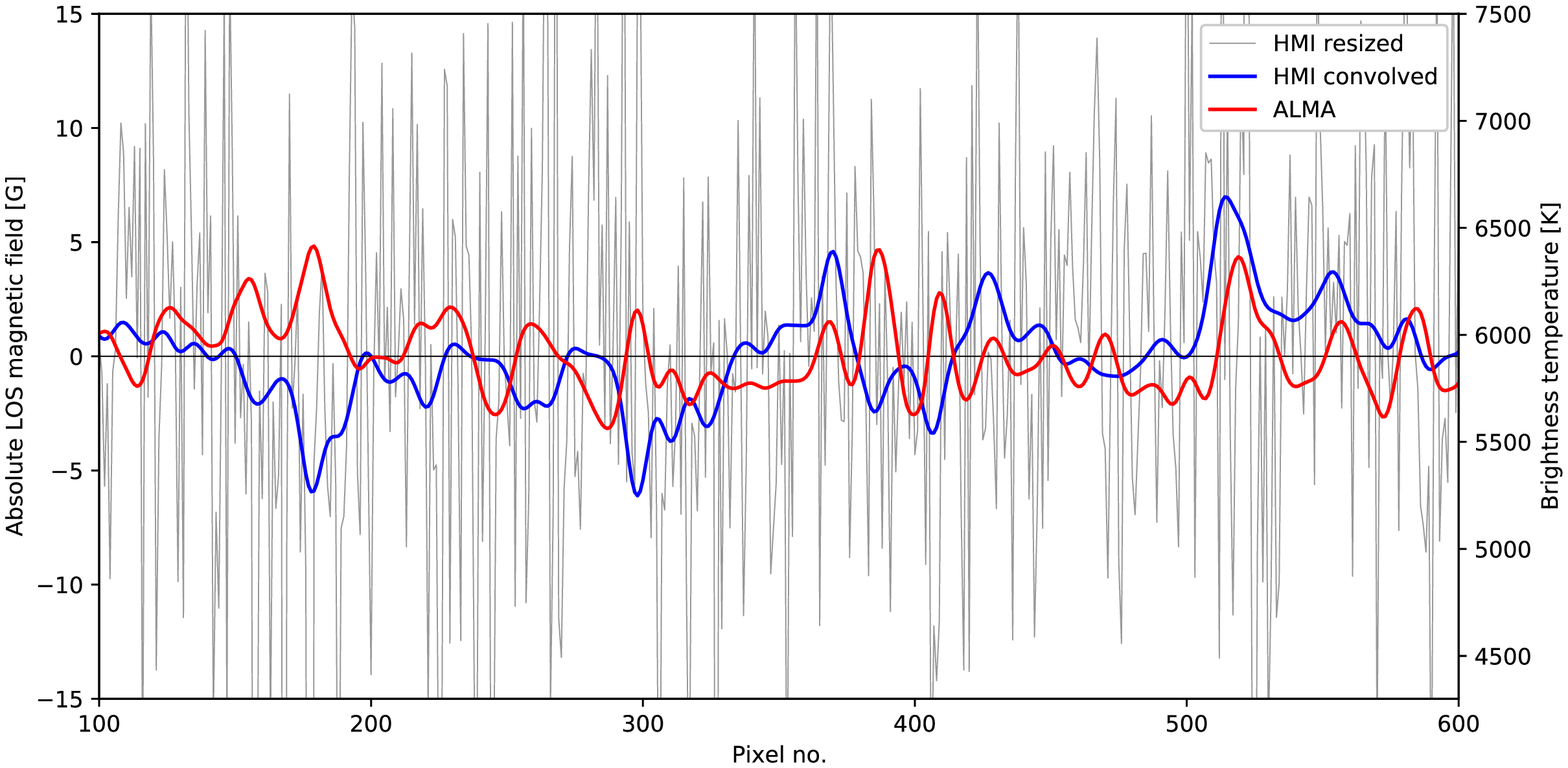,width=12.6cm}%
  \caption{Profile of the line cutting through the center of the Sun. HMI data resized to ALMA image size is shown in gray, convolved and resized HMI data is shown in blue, while ALMA intensity is shown in red. Note the different axes for HMI and ALMA data. Black horizontal line denotes LOS magnetic field strength of zero Gauss and quiet Sun brightness temperature for ALMA band 6 of 5900 K.}
  \label{fig:line_profiles}
\end{figure}

The main idea is to compare the ALMA brightness temperature with the HMI LOS magnetic field strength and look for possible correlation between the two. First, the single line profiles from ALMA and HMI, cutting through the center of the Sun, are compared. Then, the whole images are compared on a pixel-by-pixel basis.

\section{Results and discussion}

Profiles of the line cutting through the center of the Sun are shown in Fig. \ref{fig:line_profiles}. The gray line denotes resized only HMI data, the blue line denotes convolved and resized HMI data, while the red line represents ALMA data. A horizontal black line of zero Gauss is added to mark the polarity reversal of the HMI data, corresponding also to the quiet Sun level of 5900 K. It can be seen that ALMA brightness temperature ($T_b$) is correlated pretty well with the positive, and anti-correlated with the negative values of the convolved HMI LOS field strength ($B_{LOS}$). This (anti)correlation is not that obvious when using the non-convolved HMI data, implicating the importance of correctly preprocessing the data and performing the comparison at the same spatial scale and resolution.

\begin{figure}
  \centering
   \epsfig{file=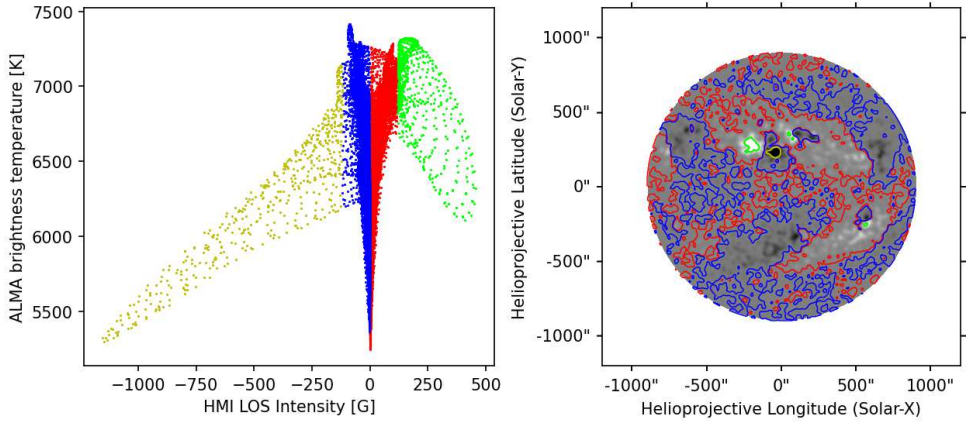,width=12.6cm}%
  \caption{ALMA 248 GHz brightness temperature compared with HMI LOS magnetic field $B_{LOS}$ (left). Data set was divided into groups based on $B_{LOS}$ and marked with different colors, and the corresponding regions are shown in the HMI image on the right. Tbe two wings, left (yellow) and right (green), come from two sunspots of opposite polarity, as denoted in the right panel.}
  \label{fig:tb_hmi_overview}
\end{figure} 

\begin{figure}
  \centering
   \epsfig{file=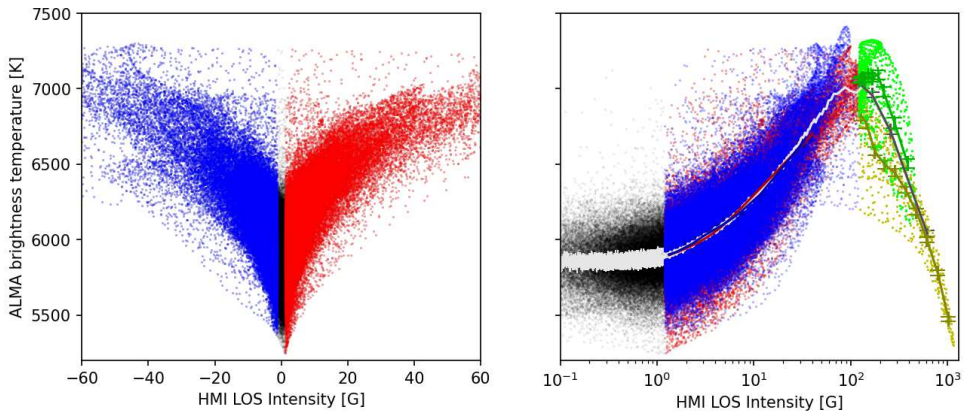,width=12.6cm}%
  \caption{ALMA brightness temperature and HMI LOS magnetic field strength comparison as in Fig. \ref{fig:tb_hmi_overview}, but zoomed to a region of $\pm60$ G (left) and with a logarithmic scale to show the dependence in the region of low magnetic field values (right). In the right panel, absolute LOS values are shown. The same colors are used as in Fig. \ref{fig:tb_hmi_overview} Colored lines denote binned values of data subgroups while white/black line is binned over the whole data set. Bins have equal number of data in each bin.}
  \label{fig:tb_hmi_log}
\end{figure}

The comparison of the whole disk $T_b$ and $B_{LOS}$ values is shown in the left panel of Fig. \ref{fig:tb_hmi_overview}. Only pixels inside the solar disk with a cutoff radius of 300 pixels (900 arcsec) were used. The apparent photospheric radius of the Sun ($6.96\times10^{8}$ m) had a value of 975.3 arcsec or 325 pixels.

The result is an interesting but complicated shape that suggests that there might be several $T_b$ -- $B_{LOS}$ correlations hidden inside the graph. To investigate this further, pixels were divided into five subgroups denoting different regions of the graph, based on their HMI $B_{LOS}$ value: $B_{LOS}<-120$ G (yellow), $-120<B_{LOS}<-1.2$ G (blue), $-1.2<B_{LOS}<1.2$ G (black), $1.2<B_{LOS}<120$ G (red), and $B_{LOS}>120$ G (green). Not surprisingly, these subgroups actually trace different regions on the Sun, namely, positive (red) and negative (blue) polarity active regions and chromospheric/magnetic network, positive (green) and negative (yellow) polarity sunspots, and the remaining pixels come from the quiet Sun/inter-network regions.

Interesting is a negative correlation between $T_b$  and absolute value of $B_{LOS}$ within both the positive and negative polarity sunspot; higher absolute values of $B_{LOS}$ correspond to lower $T_b$. The correlation is not perfect, there is some structure most probably coming from systematic errors like the antenna scanning pattern, which is still present in the ALMA data as small regular variations of the brightness temperature. The ALMA Solar Development Team is actively working to correct these systematic errors and other artifacts still present in many solar ALMA single-dish data. 

On the other hand, active regions suggest a positive correlation where higher absolute $B_{LOS}$ values correspond to higher $T_b$. This is visible as a narrow V shape in the left panel of Fig. \ref{fig:tb_hmi_overview}. This V shape is more obvious in a zoomed-in image in the left panel of Fig. \ref{fig:tb_hmi_log}, indicating a clear link between the $B_{LOS}$ and $T_b$ in the active solar regions. However, a small asymmetry between positive and negative active regions visible in Fig. \ref{fig:tb_hmi_overview} (red and blue stripe), is not visible in the zoomed-in Fig. \ref{fig:tb_hmi_log}.

To check if there is a correlation in quiet Sun regions as well, a logarithmic scale for $B_{LOS}$ is used in the right panel of Fig. \ref{fig:tb_hmi_log}, since average magnetic field strength is low in these regions. They are denoted with black pixels. It can be seen that these regions do not correlate with $B_{LOS}$, rather they have more or less a constant value of $T_b \approx 5900$  K.

To explore further these interesting relationships, we binned the whole data set into bins with equal number of data points. The result is shown as white and black line in the right panel of Fig. \ref{fig:tb_hmi_log}. Each subgroup was also binned and shown with the line of the same but darker color, with 1-$\sigma$ values denoted by crosses at bin centers. Bins just confirm previous conclusions: regions of $| B_{LOS} |<1.2$ G do not show any correlation, regions of $| B_{LOS}|$ between 1.2 and 120 G show a positive correlation, while regions with $| B_{LOS}|$ above $120$ G show a negative correlation.
We calculated correlation coefficients for the five subgroups of the data and got 0.735 (0.736) for positive (negative) polarity active regions, -0.916 (-0.619) for the negative (positive) polarity sunspot, and 0.074 for the quiet Sun region.

There is a significant scatter in the presented graphs with possible hints of a substructure (Figs. \ref{fig:tb_hmi_overview} and \ref{fig:tb_hmi_log}). One reason, already mentioned, is that ALMA maps are not yet perfectly calibrated and some systematic errors are still present in the data, such as variability in the brightness temperature, scan pattern visibility, "spokes" on the limb, etc. Possible causes include variable receiver gain, atmospheric conditions, antenna response, and even dynamic solar conditions, since the effective exposure time was more than 1 min without calibration scans. Moreover, by convolving the HMI data to the resolution of ALMA, averaged magnetic field values are decreased, as can be seen in Fig. \ref{fig:line_profiles}. The analysis only included LOS magnetic field without information of the field direction which introduces geometrical effects which are more pronounced near the limb. 

Despite all these effects which were not taken into account, the correlation is definitely present. In the case of sunspots, the correlation is close to linear, while active regions show pronounced nonlinearity. A similar negative correlation for sunspots was observed by \citet{Korzhavin2010} (see their Fig. 8). They used RATAN-600 radiotelescope at 6--18 GHz and assumed that the radiation is coming from a gyroresonance layer above sunspots with the brightness temperature equal to the kinetic temperature of electrons which depends on the magnetic field strength and the observed frequency.

The apparent small asymmetry between the positive and negative active regions visible in Fig. \ref{fig:tb_hmi_overview} (opposite polarity sunspots also show different behavior) might be the result of different properties of the ordinary and extraordinary wave modes, coming preferentially from different polarity regions, but it is highly uncertain.

We examine two possible radiation mechanisms to explain the observed correlation. One is gyroresonance, where non-relativistic electrons spiral along magnetic field lines, radiating along the way. The absorption coefficient is strongest at harmonic/resonant frequencies of the gyrofrequency of the electrons and quickly decreases elsewhere \citep{Nindos2020review}. Also, the contribution of higher resonant modes is negligible \citep{Brajsa2009}. For the ALMA band 6 frequency of 248 GHz, a field of $\approx 30$ kG is needed for gyroresonance emission at 3\textsuperscript{rd} harmonic, which is an order of magnitude higher than the observed field in sunspots. This leaves us with the free-free emission as the most probable explanation, where radiation is a result of free electrons interacting with ions in the plasma. In this case, magnetic field can change the local plasma properties, as is the case in sunspots where it inhibits convection. In active regions, the main source of millimeter free--free opacity are plages, chromospheric regions with higher temperature and density than the surrounding quiet Sun regions. The free--free millimeter intensity is approximately proportional to the electron temperature, so an ALMA image practically maps the temperature variations of the chromosphere.

\section{Summary and conclusion}
We analyzed the correlation between the photopsheric LOS magnetic field and the ALMA brightness temperature by comparing the HMI magnetogram and the ALMA single-dish image of the Sun. The analysis was performed on five subgroups of data based on their magnetic field strength. No correlation was observed for the quiet Sun regions, strong negative correlation was found for both positive and negative polarity sunspots, while network and active regions show a positive correlation. The character of the observed correlation seems to change at $| B_{LOS}|\approx 1$ G (no correlation to positive one) and $\approx 120$ G (positive to negative). The probable explanation is that observed millimeter radiation is mostly coming from the thermal bremsstrahlung of solar plasma, where the magnetic field affects the temperature and density of the plasma and thereby modulates the radiated intensity. Measured field strengths are far too weak for the gyroresonance to contribute significantly to the free-free opacity.  

In this comparison between the LOS magnetic field and the brightness temperature, we have removed the limb brightening profile from the ALMA image, which is a questionable step because it removed the $T_b$ dependency with the line-of-sight angle. However, we also used a radial limit for the data, so regions near the limb with a significant brightening would be removed anyway from the analysis, even without the limb-brightening correction. The effect of limb brightening on the results will be a topic of further research on the subject.

\section*{Acknowledgements} 
This work has been supported by the Croatian Science Foundation under the project 7549 "Millimeter and submillimeter observations of the solar chromosphere with ALMA".
This paper makes use of the following ALMA data: ADS/JAO.ALMA\#2011.0.00020.SV. ALMA is a partnership of ESO (representing its member states), NSF (USA) and NINS (Japan), together with NRC (Canada), MOST and ASIAA (Taiwan), and KASI (Republic of Korea), in cooperation with the Republic of Chile. The Joint ALMA Observatory is operated by ESO, AUI/NRAO and NAOJ. We acknowledge the use of the ALMA Solar Ephemeris Generator \citep{skokic2019seg} for preparation of solar ALMA observations.
This research used version 2.1.2 \citep{sunpy_ver_212} of the SunPy open source software package \citep{sunpy2020}. HMI image is courtesy of NASA/SDO and the AIA, EVE, and HMI science teams. 

\medskip
\medskip
\medskip
\medskip

\bibliographystyle{ceab}
\bibliography{hmi_full}

\end{document}